\documentclass{emulateapj}

\usepackage{graphicx}

\def\etal{{et~al.\,}}

\def\sles{\lower2pt\hbox{$\buildrel {\scriptstyle <}
   \over {\scriptstyle\sim}$}}
\def\sgreat{\lower2pt\hbox{$\buildrel {\scriptstyle >}
   \over {\scriptstyle\sim}$}}

\slugcomment{Accepted to Ap.J. May 2, 2008}

\begin{document}

\title{Optical Albedo Theory of Strongly-Irradiated Giant Planets: The Case of HD 209458b} 

\author{A. Burrows\altaffilmark{1,2}, L. Ibgui\altaffilmark{2} \& I. Hubeny\altaffilmark{2}} 

\altaffiltext{1}{Department of Astrophysical Sciences, Princeton University, Princeton, NJ 08544;
burrows@astro.princeton.edu}

\altaffiltext{2}{Department of Astronomy and Steward Observatory, 
                 The University of Arizona, Tucson, AZ \ 85721;
                 burrows@as.arizona.edu, ibgui@as.arizona.edu, hubeny@aegis.as.arizona.edu}

\begin{abstract}

We calculate a new suite of albedo models for 
close-in extrasolar giant planets and compare with the recent stringent 
upper limit for HD 209458b of Rowe et al. using MOST.  We find that all 
models without scattering clouds are consistent with this optical limit. 
We explore the dependence on wavelength and waveband, metallicity, the degree of heat redistribution, and 
the possible presence of thermal inversions and find a rich diversity of behaviors.  
Measurements of transiting extrasolar giant planets (EGPs) at short
wavelengths by MOST, Kepler, and CoRoT, as well as by proposed dedicated multi-band 
missions, can complement measurements in the near- and mid-IR using {\it Spitzer} 
and JWST.  Collectively, such measurements can help determine metallicity, compositions, 
atmospheric temperatures, and the cause of thermal inversions (when they arise)
for EGPs with a broad range of radii, masses, degrees of stellar insolation, 
and ages. With this paper, we reappraise and highlight the diagnostic 
potential of albedo measurements of hot EGPs shortward of $\sim$1.3 $\mu$m. 

\end{abstract}


\section{Introduction}
\label{intro}

Optical albedo measurements of solar system objects such as planets, asteroids, and moons
have a strong tradition in planetary science.  The associated reflectances and their wavelength
dependence have been used to determine surface compositions and, in concert with 
infrared (IR) photometric measurements and orbital distances, can be used to determine  
radii.  Orbital measurements can sometimes reveal masses and,
given the radius, an object's average density can be calculated.  This density
then points to its bulk composition.  In this way, optical albedo studies have often been
central to constraining the interior and surface properties of solar-system bodies. 

With the discovery of extrasolar giant planets (EGPs), this same program of sensing
and characterization can be envisioned and numerous pilot programs to measure their
optical albedos have been inaugurated (Collier-Cameron et al. 2002; Leigh et 
al. 2003; Leigh, Collier-Cameron, \& Guillot 2003; Rowe et al. 2006). However, unless the EGP can be imaged 
separately from its primary star, one must contend with the severe planet$-$star flux
ratios in the optical.  At the distance of Jupiter from its primary, this ratio is $\sim$$10^{-9}$. 
Even at the distance of the closest transiting EGPs ($\sim$0.02 A.U.),
the optical planet$-$star flux ratio is around $10^{-5}$ (Sudarsky, Burrows, \& Pinto 2000; Marley et al. 1999).
These small numbers are the primary reason direct measurements in the near- to mid-infrared
using Spitzer have been more productive of late in the study of close-in and 
transiting EGPs. The corresponding ratios of the absorbed optical radiation 
(reprocessed by the planet into the infrared) to the stellar IR can be as high as $10^{-4}$ 
to 10$^{-2}$. With such favorable contrasts, and using the {\it Spitzer} IRAC 
and MIPS IR photometers and the IRS spectrometer, planets outside the solar system 
have been directly measured and characterized for the first time (Charbonneau et al. 2005; 
Deming et al. 2005,2006,2007; Harrington et al. 2006; Harrington et al. 2007; 
Knutson et al. 2007; Grillmair et al. 2007; Richardson et al. 2007; Swain et al. 2008). Researchers 
have measured atmospheric temperatures, thermal inversions, and the possible presence
of water, methane and carbon monoxide.  They have also discerned atmospheric features
and inferred day-night climatological contrasts (Harrington et al. 2006; Knutson et al. 2007).

Nevertheless, the optical offers opportunities for measurements that can complement those in 
the IR to provide firmer constraints on the atmospheric properties of close-in EGPs. The albedo in 
the blue can reveal Rayleigh scattering off atmospheric H$_2$ and He to determine 
particle densities.  Measurements in the resonance lines of the alkali metals Na 
($\sim$0.589 $\mu$m) and K ($\sim$0.77 $\mu$m) directly test for their presence 
and are a counter to transit radius measurements that have already indicated
the presence of Na in HD 209458b (Charbonneau et al. 2002).  Importantly, albedo/reflection 
measurements probe different regions of the atmosphere than transit measurements. 

The depth of the water feature around $\sim$0.94 $\mu$m addresses both the abundance of oxygen
and of water.  In these hot atmospheres, oxygen can be mostly in CO and the CO/H$_2$O ratio 
might someday provide a signature of non-equilibrium chemistry and overall metallicity. 
The latter addresses planetary origins and formation. 

The thermal inversions recently invoked to explain the photometric reversals and flux 
enhancements seen in the IRAC, IRS, and/or MIPS data for HD 209458b (Hubeny, 
Burrows, \& Sudarsky 2003; Knutson et al. 2008; Burrows et al. 2007) and for HD 149026b 
(Harrington et al. 2007; Burrows, Budaj, \& Hubeny 2008; Fortney et al. 2006) might be due to an 
``extra absorber" in the optical. Such an absorber would have a strong impact on the 
optical reflection spectrum.  If it were TiO, VO, and/or a chemical product of photolysis, 
such as polyacetylenes or ``tholins," distinctive absorption features would in principle be 
observable\footnote{Note that Jupiter's albedo spectrum in the blue is lowered by about a 
factor of two shortward of $\sim$0.5 $\mu$m due to the presence of non-equilibrium upper-atmosphere absorbing 
species of unknown origin.}. If the temperature inversion were very strong, elevating 
the temperature of the upper atmosphere above the ``plateau" value at depth, emission in the optical might trump 
absorption (Hubeny, Burrows, \& Sudarsky 2003; Seager \& L\'{o}pez-Morales 2007). On the other hand, 
if the inversion is due not to an optical absorber, but to a mechanical heat source like the 
breaking of inertial or tidal waves, the resultant absence of a distinctive 
optical absorption feature in the albedo spectrum would be telling. These speculations all
serve to emphasize the fact that theory is still in its earliest formative stages, for 
though inversions in transiting EGPs were anticipated, a thermal inversion specifically in 
HD 209458b, and the magnitude of that inversion, were not.

In addition, there could be thick cloud layers of silicates and/or
iron.  Such clouds are seen in L dwarf atmospheres and might 
emerge in an irradiated planet's atmosphere when their condensation 
curves intersect the planet's $T/P$ profile.  This could happen
for the most severely irradiated EGPs. If these clouds form, they
do so first near the sub-stellar point and and equatorial regions.  Reflecting clouds
could increase the optical albedo by a factor of $\sim$2 to $\sim$5.

Moreover, the redistribution of heat by zonal winds from the dayside to
the nightside lowers the dayside temperatures and, thereby, alters
the molecular composition.  If we can measure this effect, parametrized
by P$_n$ ($0 <$P$_n$$< 0.5$) (Burrows et al. 2005, 2006, 2007), we might have a handle
on the planet's global climate.  In the optical and very-near-IR, the
variation with P$_n$ of the planet's flux and albedo can be a factor 
of 50\%, depending upon wavelength.  Note that the magnitude of the
dependence on P$_n$ of the planet$-$star flux ratio is no weaker 
in the optical than in the mid-IR probed by {\it Spitzer}.

Therefore, due to the possible features, effects, and constituents mentioned above, the
monochromatic optical albedos from $\sim$0.35 to $\sim$1.0 $\mu$m for the family of known transiting
EGPs could vary by about an order of magnitude.  Moreover, due to physical 
differences between the planets, the slopes of the albedo spectra at various wavelengths 
and the contrasts in and out of various bands can differ by many factors.  The flux at the sub-stellar
point (hereafter ``sub-stellar flux") on the known transiting planets ranges by $\sim$two orders of magnitude
from $0.03 \times 10^9$ ergs cm$^{-2}$ s$^{-1}$ for GJ 436b to $5.9 \times 10^9$ ergs cm$^{-2}$ s$^{-1}$ for OGLE-TR-56b
and the stellar types of the parents range from M, through K and G, to F.
Therefore, the irradiation spectra vary substantially and this variety should translate
into a variety of albedo spectra. In sum, optical albedo measurements complement near- 
and mid-IR measurements to more strongly constrain the physical and chemical properties
of tranisting EGP atmospheres than either can do alone.

Among those transiting EGPs that orbit bright and/or nearby stars, HD 189733b,
HD209458b, TrES-1, HAT-P-2b, GJ 436b, XO-2b, HAT-P-1, HD149026b, and TrES-4 collectively
constitute a heterogeneous class of objects that must have disparate
albedo and flux signatures that will enable a fruitful photometric study of the
atmospheres of close-in EGPs.  For a non-transiting close-in EGP (such
as 51 Peg b, $\upsilon$ And b, or $\tau$ Boo b), the ambiguity in the orbital
inclination and radius renders interpretation more difficult, but no less interesting.

Recently, using the MOST (Microvariability \& Oscillations of STars) micro-satellite 
(Walker et al. 2003; Matthews et al. 2004), Rowe et al. (2007) obtained a stringent
1-$\sigma$ upper bound of 8.3\% on the integrated optical albedo of HD 209458b\footnote{Rowe et al.
achieve a photometric accuracy better than $10^{-5}$.}.  The actual
measurement is 3.8\% $\pm$ 4.5\%.  More data on other transiting EGPs are sure to follow
and this, along with the new HD 209458b constraint, make it a good time to address
anew the theory of EGP albedos. In this paper, we compare the Rowe et al. number
with a representative collection of new albedo models.  Though we focus on HD 209458b,
our subsidiary goal is to reignite the discussion concerning the influences 
on EGP albedos of variations in the physical properties of their 
atmospheres.

\section{Albedo Models and Comparision with the HD 209458b Data}
\label{models}

The spherical albedo is the fraction of the stellar light at a given wavelength 
reflected into any and all angles.  The Bond albedo is the frequency or wavelength integral of the 
spherical albedo, weighted by the incident stellar spectrum, and is a measure
of the heat reflected.  The Bond albedo is the quantity one uses to determine
the planet's total energy budget.  However, it is the geometric albedo (A$_g$) that one measures
when detecting the planet's flux at the Earth.  Hence, this albedo 
is the focus of this report and it can simply be defined using the formula:

\begin{equation}
{\rm F}_p/{\rm F}_*  = {\rm A}_g \Bigl(\frac{{\rm R}_{\rm p}}{{\rm A}}\Bigr)^2 \, ,
\label{define}
\end{equation}
where F$_p$/F$_*$ is the planet$-$star flux ratio at full face, which for transiting
planets is at secondary eclipse, R$_{\rm p}$ is the planet radius, and A is the semi-major axis 
of the planet's orbit, assumed hereafter to be circular.  Of course, the planet$-$star flux ratio
varies with phase angle via the phase function (Burrows, Sudarsky, \& Hubeny 2004; Sudarsky, 
Burrows, \& Hubeny 2005), but at superior conjunction the phase function is set to one, 
thereby fixing the definition of A$_g$.  Henceforth, we will not be concerned with the phase dependence 
and will focus on the full-face quantities A$_g$ and F$_p$/F$_*$ and their wavelength dependence.

Implicitly, the definition in eq. (\ref{define}) acknowledges
that at the high atmospheric temperatures of close-in planets  
the planet itself can be self-luminous in the near- and mid-IR and that 
A$_g$ at these wavelengths can exceed one.  This possibility is not at all anomalous, since
in determining an albedo one is taking the ratio of planet flux to stellar
flux, at the same wavelength.  In the IR, the planet's flux may be due more to
the absorption and re-emission of the star's optical flux than to its IR flux and when
the star's IR flux is the divisor (as in eq. \ref{define}), the resulting A$_g$ can easily exceed one without
violating energy conservation.  Hence, the anomaly, if anomaly there is, is merely one of 
expectations colored by solar-system experience, for which the bodies in question are much 
cooler than the Sun and the concept of optical albedo is sensibly tied to reflection. 
However, for close-in EGPs, the fluxes are more and more thermal longward of $\sim$0.7 $\mu$m. 
Nevertheless, incorporating the stellar flux as a boundary condition and solving the 
equations of radiative transfer for all wavelengths self-consistently, yields 
values of F$_p$/F$_*$, from which A$_g$ as a function of wavelength can be  
derived using eq. (\ref{define}).  

To produce our models for F$_p$/F$_*$ and A$_g$,
we employ the self-consistent atmosphere code and solution techniques 
described in Hubeny \& Lanz (1995), Burrows, Hubeny, \& Sudarsky (2005), 
Burrows, Sudarsky, \& Hubeny (2006), and Burrows, Budaj, \& Hubeny (2008), 
to which the reader is referred for details.  In Burrows, Budaj, \& Hubeny (2008),
we also describe how we now handle heat redistribution for a given 
P$_n$ and how we derive the flux at secondary eclipse.  In Sharp \& Burrows (2007), 
we review our opacity libraries and chemical abundance calculations.  The latter
are done in the context of chemical equilibrium. Using these tools and 
databases, we have generated spectral models for HD 209458b at superior 
conjunction as a function of metallicity and P$_n$.  We have also explored the
dependence on a possible high-altitude ``extra absorber" in the optical, as inferred for HD 209458b
by Burrows et al. (2007) using the IRAC data of Knutson et al. (2008). Though the models
range from 0.3 $\mu$m to 300 $\mu$m, we highlight in this paper only the results
shortward of 1.3 $\mu$m.  

\begin{figure}
\centerline{
\includegraphics[width=16.cm,angle=0,clip=]{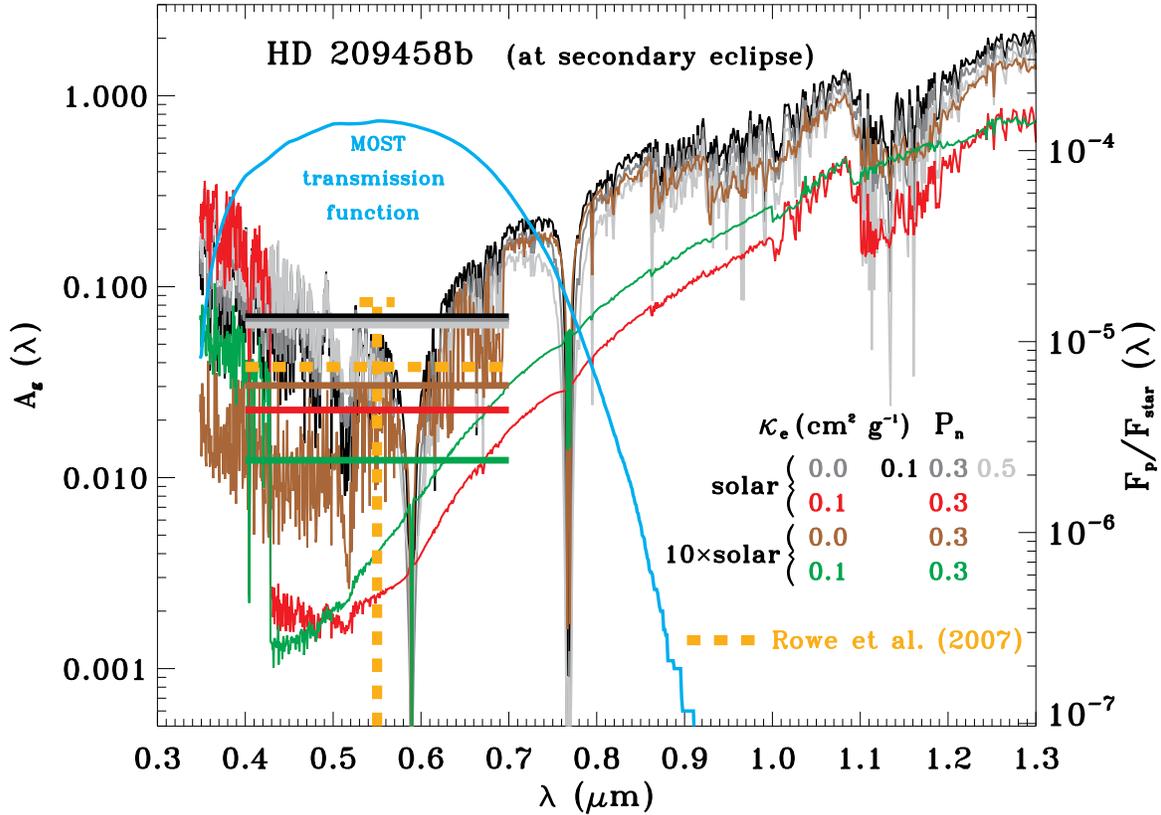}}
\caption{The logarithm base 10 of theoretical albedos (left axis) and 
F$_p$/F$_*$ ratios (right axis) as a function of wavelength (in microns) 
from 0.35 $\mu$m to 1.3 $\mu$m for a collection of models for HD 209458b.  Here 
F$_{\rm star}$ = F$_*$. Each model spectrum is for a given representative combination 
of P$_n$ (either 0.1, 0.3, or 0.5), atmospheric metallicity (either solar or 10$\times$solar), 
and ``extra absorber" opacity ($\kappa_{\rm e}$, either 0.0 or 0.1 cm$^2$/$g$). The legend at the 
bottom right provides the matrix of models and the colors of the corresponding lines.  The 
prescription for incorporating the extra absorber, when present, is described in Burrows, 
Budaj, \& Hubeny (2008).  $\kappa_{\rm e}$ is a phenomenological parameter with which 
to numerically generate a thermal inversion at low pressures.  Such an inversion is inferred 
for HD 209458b by Burrows et al. (2007) from an analysis of its IRAC data (Knutson 
et al. 2008) and may be due to photolytic products, TiO/VO, or some other origin.  
Superposed in blue is the MOST transmission function. Since its transmission 
function is so broad, ranging from $\sim$0.35 $\mu$m to 0.8 $\mu$m, we provide with 
horizontal lines the corresponding predicted average albedos as they would be measured 
by MOST.  The Rowe et al. (2007) data (3.8\% $\pm$ 4.5\%) are superposed in orange (dashed),
with the topmost extent indicating the 1-$\sigma$ upper limit (8.3\%). See the text for 
a discussion of the models and their implications. 
}
\label{fig1}
\end{figure}

Figure \ref{fig1} portrays the albedo spectra for a variety of atmospheric
models for HD 209458b.  Superposed are the Rowe et al. (2007) upper limit
(8.3\% [1-$\sigma$] $\equiv$ 3.8\%$\pm$4.5\%), the MOST transmission curve,
and the model bandpass-averaged albedos.  Table 1 depicts these averaged albedos,
as well as the corresponding numbers for a few models of HD 189733b for comparison.
The model spectra are calculated for a given triplet of P$_n$ (0.1, 0.3, or 0.5), 
atmospheric metallicity (either solar or 10$\times$solar), and 
$\kappa_{\rm e}$ (0.0 or 0.1 cm$^2$/$g$). $\kappa_{\rm e}$ is the opacity 
of a putative extra absorber which might be necessary to generate the 
thermal inversion at altitude inferred from the IRAC data for HD 209458b 
(Knutson et al. 2008). We distribute this absorber everywhere 
in the atmosphere at pressures lower than $\sim$0.014 bars 
%
%
and in the wavelength interval  
0.43 $\mu$m $< \lambda <$ 1.0 $\mu$m.  This prescription should be viewed 
as useful, but simplistic.  To fit the HD209458b data, Burrows et al.
(2007) employed a constant value of $\kappa_{\rm e}$ = 0.1 cm$^2$/$g$.  

Figure \ref{fig1} incorporates our major results and manifests a number of features.   
The general trend is that the albedo decreases to shorter wavelengths
until Rayleigh scattering off H$_2$ and He reverses this.  This can occur near $\sim$0.5 $\mu$m
and can be responsible for an increase in A$_g$ of as much as an order of magnitude
from 0.55 $\mu$m to $\sim$0.30 $\mu$m.  At its trough, A$_g$ can be as low as (or lower than)
$\sim$1\%, but without an extra broad-band absorber as we have modeled it here
it is only in the relatively narrow alkali metal features that A$_g$
is severely low.  Otherwise, the trough is around $\sim$1\% to $\sim$5\%.  However, this can correspond
to values of F$_p$/F$_*$ less than 10$^{-5}$ and such numbers may be inaccessible from the ground
for the foreseeable future.  However, due to the high temperatures ($\sim$1500$-$2000 K) 
in HD 209458b's atmosphere (Burrows, Budaj, \& Hubeny 2007), A$_g$ exceeds 1.0 
longward of $\sim$1.0 $\mu$m to $\sim$ 1.1 $\mu$m.   In fact, using the strict definition
in eq. (\ref{define}), in the mid-IR A$_g$ exceeds 10.  Note that for most of our models
longward of $\sim$0.7 $\mu$m A$_g$ is above 0.1 (10\%).  At $\sim$1.0 $\mu$m, it can be
near $\sim$60\%.  Therefore, longward and shortward of the $V$ and $R$ bands, 
in the $U$, $B$, $Z$, $Y$, and $z^{\prime}$ bands, A$_g$ can be respectable, even high.  
Such high values may explain the very tentative 1-$\sigma$ lower limit 
to A$_g$ in the $B$ band of $\sim$0.14 estimated by Berdyugina et al. 
(2007) for HD 189733b (see also Table 1).

However, when integrating over the MOST passband, our models, for the variety of
values of P$_n$ and metallicity, have low predicted average albedos, varying 
from $\sim$0.017 to $\sim$0.07 (Table 1).  These are quite ``black."
In fact, all our models are consistent with the low Rowe et al. 1-$\sigma$ upper limit 
of 8.3\%.   Reflecting clouds in the atmospheres of solar-system bodies,
and in previous theoretical EGP models with clouds (e.g., Sudarsky, Burrows, \& Pinto 2000), have 
values of A$_g$ above 0.2 (20\%), usually near $\sim$30-40\%.  Such numbers are inconsistent
with the Rowe et al. measurement.  Therefore, a straightforward conclusion is that 
there are no reflecting clouds in the atmosphere of HD 209458b.  Thin hazes are 
possible, but we estimate their optical depth in the MOST bandpass can be no larger 
than $\sim$0.1 for high single-scattering albedos ($\Sigma$ = $\kappa_s$/($\kappa_s + \kappa_a$)). 
This conclusion may challenge some otherwise reasonable solutions (Fortney et al. 2003) 
to the anomaly in the magnitude of the transit radius in the Na-D lines measured 
by Charbonneau et al. (2002).

The Rowe et al. (2007) data provide only one number and the MOST transmission curve is
very broad.  As a result, we can derive little from these data about the atmosphere
of HD 209458b.  However, the variety of models depicted in Fig. \ref{fig1} and Table 1
allow us to make general predictions concerning the diagnostic potential of future albedo
spectrum measurements from $\sim$0.3 $\mu$m to 1.3 $\mu$m and to discuss the dependence
of A$_g$ on the physical attributes of the atmospheres of close-in EGPs.  

The relative behavior in Fig. \ref{fig1} of the solar metallicity models at 
$\kappa_{\rm e}$ = 0 (black and two shades of gray) as a function of P$_n$ 
demonstrates that small values of P$_n$, which yield the hottest dayside 
temperatures, result in higher values of A$_g$ longward of $\sim$0.65 $\mu$m.  
A$_g$ and F$_p$/F$_*$ increase by $\sim$30-50\% as P$_n$ shifts from 0.5 to 0.1.
This is as expected, given the thermal character of the longer-wavelength fluxes.    
However, hotter atmospheric temperatures result in higher absorption opacities.
At the shorter wavelengths, where Rayleigh scattering is manifest, these higher
absorption opacities translate into lower scattering albedos.  As Sudarsky, Burrows, 
and Pinto (2000) and van de Hulst (1974) showed, A$_g$ decreases quickly with decreasing
single-scattering albedo, as much as a factor of two for a $\sim$10\% decrease in $\Sigma$.
We see this behavior in Fig. \ref{fig1} between 0.4 and 0.5 $\mu$m, where the black curve
(P$_n$ = 0.1) is almost a factor of two lower than the light gray curve (P$_n$ = 0.5),
even though the MOST averages are comparable.  Hence, higher values of the heat 
redistribution parameter result in higher (lower) values of A$_g$ at short (long)  
wavelengths.

Comparing the brown curve (10$\times$ solar, P$_n$ = 0.3) with the dark gray curve
(solar, P$_n$ = 0.3) allows one to gauge the metallicity dependence.  However, shortward
of $\sim$1.3 $\mu$m, the metallicity effect can be subtle.  At wavelengths longer than $\sim$0.9 $\mu$m,
higher metallicity results in slightly lower A$_g$ and F$_p$/F$_*$, but only by $\sim$10\%.  Between 
$\sim$0.85 $\mu$m and $\sim$0.55 $\mu$m, the A$_g$s are very similar.  However, below $\sim$0.5 $\mu$m,
A$_g$s are much lower for 10$\times$solar metallicity.  The higher metallicity
results in much higher absorption cross sections and, hence, lower $\Sigma$.   
In a spectral region where Rayleigh scattering, not thermal emission, is important, 
as described above this translates into a much lower value of A$_g$.  This fact is demonstrated 
in Table 1 by comparing the $B$ band albedos.  

Note that the transmission curves of MOST, Kepler (Borucki et al. 
2003; Koch et al. 2004; D. Koch, private communication), 
and CoRoT\footnote{The transmission curve for CoRoT can be found
at \url{http://corotsol.obspm.fr/web-instrum/payload.param/}.} penetrate progressively more and 
more into the red. For MOST, the wavelength on the red side at which transmission is 10\% of peak  
is $\sim$0.75 $\mu$m, for Kepler it is $\sim$0.9 $\mu$m, and for CoRoT it is $\sim$1.0 $\mu$m.
As indicated in Table 1, the different predicted albedos for these different satellites
may be crudely diagnostic of the generic wavelength dependence, with CoRoT
sampling the longer wavelengths at which A$_g$ is generally higher. However,
since CoRoT, MOST, and Kepler can't observe the same stars, until
there is a dedicated multi-band albedo mission, we will have to make due 
with minimal constraints, however precious, on any particular transiting EGP albedo spectrum.  

Water absorption features reside at $\sim$0.92-1.0 $\mu$m and $\sim$1.1-1.2 $\mu$m and
are clearly seen in Fig. \ref{fig1}.  Their depths can be from $\sim$50\% to a factor of two
and are clear signatures of water's presence.  Nearly saturated, the metallicity and abundance dependence is modest,
with the 10$\times$solar model lower by $\sim$20-40\% than the solar model.  Hence, measurements 
in and out of these features and of the absolute A$_g$s would identify water's presence and might 
constrain its abundance.

The red (solar) and green (10$\times$solar) curves in Fig. \ref{fig1} depict two realizations 
of a model with an  ``extra absorber" ($\kappa_{\rm e}$ = 0.1 cm$^2$/g) in the optical 
that creates a thermal inversion at altitude and explains the 
the IRAC, MIPS, and IRS data for HD 209458b (Burrows et al. 2007). While in the mid-IR these 
two models differ little ($\sim$20\%), the presence of a broad optical absorption feature
can severely suppress flux where it operates.  At solar metallicity (red), the effect on A$_g$ is
severe between its artificial blue (0.43 $\mu$m) and red (1.0 $\mu$m) edges.  However, if the metallicity
is as high as 10$\times$solar, the suppressive effect of adding an extra absorber with a given opacity is 
less (green curve). While at high metallicity the flux is generally lower in the optical than for all the models without inversions, it is
a factor of $\sles$2 times higher than for solar-metallicity models (red curve).  This inverse dependence
on metallicity is encouraging as a diagnostic, until one realizes that in an absolute sense the 
geometric albedos and flux ratios are quite low.  In addition, our models place the extra absorber
at the same pressure levels ($\sles$0.014 bars), independent of metallicity, and, depending
upon the origin of the absorber, it is likely that its physical extent will somehow be metallicity-dependent.  
Nevertheless, a metallicity dependence may be a future signature of the metallicity/extra-absorber 
combination that is not so manifest at infrared wavelengths. Clearly, knowledge of the specific absorption 
spectrum of the extra absorber, if it exists, would make our predictions more specific. 
Such would be the case if it were TiO and/or VO (Hubeny et al. 2003). 
However, the cold-trap effect (Burrows et al. 2008) suggests that TiO and VO might 
not have the requisite abundance at altitude to explain the inversion inferred 
from the HD 209458b IRAC data of Knutson et al. (2008).  This is an open question.

There are two final features to note concerning the models with an ``extra absorber" portrayed in Fig. \ref{fig1}.
The first is that at the shortest wavelengths the solar model has higher values of A$_g$ (red curve).  These higher albedos 
are a consequence of the fact that while the extra absorber heats the atmosphere at low pressures ($< 0.014$ bars),
it results in lower temperatures for the high-pressure plateau (Burrows, Budaj, \& Hubeny 2008, their Figure 1).
This is near were the far-blue and near-UV photospheres reside. Lower temperatures result in
lower absorptive opacities and, hence, higher scattering albedos.  The result is the counter-intuitive
enhancement in the albedos at the shortest wavelengths.  This is less the case for the 10$\times$solar model,
due to the correspondingly higher absorptive opacity.  The second feature is that the model with an 
extra absorber that fits the HD 209458b IRAC data (Burrows et al. 2007) predicts mid-optical albedos and corresponding planet$-$star flux ratios
that are lower, not higher, than without the absorber, mildly at odds with the suggestion of Fortney et al. (2007). 
This is because in these extra-absorber models the optical photosphere is at lower temperatures 
(and pressures) than is the temperature plateau at depth, near where the optical photosphere 
would reside without the extra optical absorber.  The upshot is a cooler emitting surface at optical wavelengths
with the extra absorber, than without, at least for the Burrows et al. (2007) models that 
currently fit the HD 209458b IRAC data. Of course, a more precise spectral model for the thermal inversion
may lead to a different conclusion, depending on the resultant $T/P$ profile. In addition, the much
hotter upper atmospheres expected for OGLE-TR-56b, TrES-4, OGLE-TR-132b, and XO-3b may indeed manifest this
albedo enhancement (Hubeny, Burrows, \& Sudarsky 2003; Seager \& L\'{o}pez-Morales 2007). However, 
our results stress caution in making qualitative predictions for planets 
with thermal inversions for the planet$-$star flux ratios in the optical.

Moreover, if the thermal inversion is due to the breaking of mechanical waves, and not an upper atmosphere
absorber in the optical, the albedo predictions could be very different.  In particular, A$_g$ in the 
mid-optical could be a factor of $\sim$5-10 higher if the inversion is not due to a extra absorber.
This difference might be a diagnostic of the wave breaking mechanism in planet atmospheres 
for which the near- and mid-IR data demand a thermal inversion.

\section{Summary}
\label{conclusions}

Focussing on HD 209458b, we have generated a new suite of albedo models for close-in extrasolar giant planets.  
We find that all models without scattering clouds fit the Rowe et al. (2007) albedo limit 
for HD 209458b and that it is no surprise that the albedo in the MOST bandpass is very low.
We have explored the effects of metallicity, of the degree of heat redistribution, and of enhanced
optical opacities at altitude that might create thermal inversions such as are inferred from the IRAC data
for HD 209458b and HD 149026b.  Measurements of hot EGPs at short wavelengths can complement measurements
in the near- and mid-IR using {\it Spitzer} and JWST to determine metallicity, water abundances,
atmospheric temperatures, and the cause of thermal inversions, when they arise.  Moreover, albedos
at short wavelengths are strongly affected by Rayleigh scattering off molecular hydrogen and helium
and their measurement can be used to determine or constrain the sum of their number densities.
Soon, Kepler and CoRoT will measure in slightly different wavebands the optical transits, the light curves, and the 
albedos of a large number of known and presently unknown transiting EGPs, thereby 
providing a wealth of new data and crude spectral diagnostics below $\sim$1.0 $\mu$m. 
These data can be compared with models calculated using the formalism we applied here for HD 209458b
to constrain atmospheric parameters of close-in giants with a broad range of gravities, degrees of stellar
insolation, abundances, and ages.  Furthermore, a number of satellites dedicated to 
albedo measurements of dozens of close-in EGPs in several bands are being proposed.
With this paper, we hope to reignite the discussion concerning 
EGP albedos and their diagnostic potential.

\acknowledgments

We thank Jaymie Matthews, Jason Rowe, Heather Knutson, Dave Charbonneau, Bill Hubbard,
Peter Goldreich, and Drew Milsom for helpful discussions and
guidance and Dave Koch and the Kepler team for providing us with an electronic
version of the Kepler transmission curve.  This study was
supported in part by NASA grants NNG04GL22G, NNX07AG80G, and NNG05GG05G
and through the NASA Astrobiology Institute under Cooperative
Agreement No. CAN-02-OSS-02 issued through the Office of Space
Science.

{}


\begin{table*}
\tiny
\begin{center}
\caption{Band-Averaged Geometric Albedos for HD 209458b and HD 189733b$^1$}
\tablewidth{17.0cm}
\begin{tabular}{ccccccccccccccc}
\hline\hline
    Planet&   & Filter     &   & MOST &  &   & Kepler &  &   & CoRoT & &  &  B Bessel &    \\
\hline
    &    [Fe/H]  &  $\kappa_{\rm e}$$\setminus$ P$_n$  $=$  & 0.1  & 0.3 & 0.5 $\mid$ & 0.1  & 0.3 & 0.5 $\mid$ & 0.1  & 0.3 & 0.5 $\mid$ & 0.1 &  0.3 & 0.5   \\
\hline
   HD 209458b &    solar  &  0.0 $\mid$ & 0.070  & 0.066 & 0.063 $\mid$ & 0.111  & 0.095 & 0.079 $\mid$ & 0.136  & 0.117 & 0.097 $\mid$ & 0.077 &  0.085 & 0.094   \\
              &           &  0.1 $\mid$ & $\cdots$  & 0.023 & $\cdots$ $\mid$ & $\cdots$  & 0.014 & $\cdots$ $\mid$ & $\cdots$  & 0.026 & $\cdots$ $\mid$ & $\cdots$ &  0.057 & $\cdots$   \\
              &    10$\times$solar  &  0.0 $\mid$ & $\cdots$  & 0.030 & $\cdots$ $\mid$ & $\cdots$  & 0.076 & $\cdots$ $\mid$ & $\cdots$  & 0.092 & $\cdots$ $\mid$ & $\cdots$ &  0.012 & $\cdots$   \\
              &           &  0.1 $\mid$ & $\cdots$  & 0.012 & $\cdots$ $\mid$ & $\cdots$  & 0.022 & $\cdots$ $\mid$ & $\cdots$  & 0.031 & $\cdots$ $\mid$ & $\cdots$ &  0.015 & $\cdots$   \\
\hline
   HD 189733b &    solar  &  0.0 $\mid$ & 0.044  & 0.044 & 0.048 $\mid$ & 0.047  & 0.045 & 0.044 $\mid$ & 0.063  & 0.059 & 0.058 $\mid$ & 0.084 &  0.088 & 0.102   \\
              &           &  0.1 $\mid$ & $\cdots$  & 0.022 & $\cdots$ $\mid$ & $\cdots$  & 0.007 & $\cdots$ $\mid$ & $\cdots$  & 0.017 & $\cdots$ $\mid$ & $\cdots$ &  0.073 & $\cdots$   \\
              &    10$\times$solar  &  0.0 $\mid$ & $\cdots$  & $\cdots$  & $\cdots$ $\mid$ & $\cdots$  & $\cdots$  & $\cdots$ $\mid$ & $\cdots$  & $\cdots$  & $\cdots$ $\mid$ & $\cdots$ & $\cdots$  & $\cdots$   \\
              &           &  0.1 $\mid$ & $\cdots$  & $\cdots$ & $\cdots$ $\mid$ & $\cdots$  & $\cdots$  & $\cdots$ $\mid$ & $\cdots$  & $\cdots$  & $\cdots$ $\mid$ & $\cdots$ & $\cdots$  & $\cdots$   \\

\hline
\end{tabular}
\tablenotetext{1}{Theoretical average geometric albedos in the MOST, Kepler, CoRoT, and B (Bessell) transmission 
bands for various of the HD 209458b models depicted in Fig. \ref{fig1}, as well as for 
HD 189733b for comparison.  For the HD 189733b model with $\kappa_{\rm e}$ = 0.1 cm$^2$/g, we also place  
the extra absorber in the atmosphere below pressures of $\sim$0.014 bars.  Each band-averaged albedo is derived by multiplying the theoretical 
wavelength-dependent albedos by the corresponding normalized transmission function and then integrating 
over wavelength. This table summarizes the metallicity, P$_n$, and $\kappa_{\rm e}$ dependences we derive 
in this study.  See the text for a discussion and Fig. \ref{fig1}.
}
\label{t1}
\end{center}
\end{table*}

\clearpage

\begin{thebibliography}{99}



















\bibitem[Berdyugina et al. 2007]{berdy} Berdyugina, S.V., Berdyugin, A.V., Fluri, D.M., \& Pirola, V. 2007, accepted to \apjl, arXiv:0712.0193



\bibitem[Borucki et al. 2003]{borucki} Borucki, W. et al. 2003, Future EUV/UV and Visible Space Astrophysics 
Missions and Instrumentation. Edited by J. Chris Blades, Oswald H. W. Siegmund. 
Proceedings of the SPIE, Volume 4854, pp. 129-140 (2003)















\bibitem[Burrows et al. 2004]{bur2004f} Burrows, A., Sudarsky, D., \& Hubeny, I.
2004 \apj, 609, 407 



\bibitem[Burrows et al. 2005]{bur2005} Burrows, A., Hubeny, I., \& Sudarsky, D.,
2005 \apj, 625, L135 

\bibitem[Burrows, Sudarsky, \& Hubeny 2006]{bur06} Burrows, A., Sudarsky, D. 
\& Hubeny, I. 2006, \apj, 650, 1140   



\bibitem[Burrows et al. 2007]{bur07} Burrows, A.,  
Hubeny, I., Budaj, J., Knutson, H.A., \& Charbonneau, D. 2007, \apj, 668, L171 

\bibitem[Burrows, Budaj, \& Hubeny 2008]{budaj} Burrows, A., Budaj, J., \& Hubeny, I. 2008, accepted to \apj (arXiv:0709.4080).





\bibitem[Charbonneau \etal 2002]{Charbonneau02} Charbonneau, D.,
Brown, T. M., Noyes, R. W., \& Gilliland, R. L. 2002, \apj, 568, 377 

\bibitem[Charbonneau \etal 2005]{char05} Charbonneau, D. \etal 2005, \apj, 626, 523






\bibitem[Collier-Cameron et al. 2002]{collier} Collier-Cameron, A., Horne, K., Penny, A., \& Leigh, C. 2002, \mnras, 339, 187 
 


\bibitem[Deming \etal 2005]{deming05} Deming, D., Seager, S., Richardson, L.J., \& Harrington, J.,
2005, Nature, 434, 740


\bibitem[Deming \etal(2006)]{deming06} Deming, D., Harrington, J.,
Seager, S., Richardson, L.R. 2006, \apj, 644, 560 

\bibitem[Deming \etal(2007)]{deming07} Deming, D., Harrington, J.,
Laughlin, G., Seager, S., Navarro, S.B., Bowman, W.C., \& Horning, K. 2007,
\apj, 667, L199 (arXiv:0707.2778) 








\bibitem[Fortney et al. 2003]{fort03} Fortney, J.J., Sudarsky, D., Hubeny, I., Cooper, C.S.,
Hubbard, W.B., Burrows, A., \& Lunine, J.I. 2003, \apj, 589, 615


\bibitem[Fortney et al. 2006]{fort2006} Fortney, J.J., Saumon, D., Marley, M.S., Lodders, K.,
\& Freedman, R.S. 2006, \apj, 642, 495  


\bibitem[Fortney et al. 2007]{fort08} Fortney, J.J., Marley, M.S., Lodders, K.,
\& Freedman, R.S. 2007, submitted to \apj (arXiv:0710.2558)







\bibitem[Grillmair et al. 2007]{grillmair} Grillmair, C.J., Charbonneau, D., Burrows, A., Armus, L.,
Stauffer, J., Meadows, V., Van Cleve, J., \& Levine, D. 2007, \apj, 658, L115






\bibitem[Harrington et al. 2006]{harrington} Harrington, J.,
Hansen, B., Luszcz, S., Seager, S., Deming, D., Menou, K.,
Cho, J., \& Richardson, L. 2006, Science, 314, 623

\bibitem[Harrington et al. 2007]{harrington07} Harrington, J., Luszcz, S., 
Seager, S., Deming, D., \& Richardson, L.J. 2007, Nature, 447, 691 









\bibitem[Hubeny \& Lanz 1995]{HubenyLanz95} Hubeny, I. \& Lanz, T. 1995,
\apj, 439, 875

\bibitem[Hubeny, Burrows, \& Sudarsky 2003]{Hubeny03} Hubeny, I., Burrows, A., \& Sudarsky, D. 2003,
\apj, 594, 1011







\bibitem[Knutson et al. 2007]{kcn07b} Knutson, H., Charbonneau, D., Allen, L.E., Fortney, J.J.,
Agol, E., Cowan, N.B., Showman, A.P., Cooper, C.S., \& Megeath, S.T.  
2007, Nature, 447, 183 

\bibitem[Knutson et al. 2008]{kcn07c} Knutson, H.A., Charbonneau, D., Allen, L.E.,
Torres, G., Burrows, A., \& Megeath, S.T. 2008, \apj, 673, 526 

\bibitem[Koch et al. 2004]{koch} Koch, D.G. et al. 2004, Optical, Infrared, and Millimeter Space 
Telescopes. Edited by Mather, J.C. Proceedings of the SPIE, Volume 5487, pp. 1491-1500









\bibitem[Leigh et al. 2003]{leigh} Leigh, C., Collier-Cameron, A., Horne, K., Penny, A., \& James, D. 2003, \mnras, 344, 1271 
 
\bibitem[Leigh, Collier-Cameron, \& Guillot 2003]{leigh1} Leigh, C., Collier-Cameron, A., \& Guillot, T. 2003, \mnras, 346, 890




\bibitem[Marley et al. 1999]{gelino} Marley, M.S.,  Gelino, C., Stephens, D., Lunine, J.I., \& Freedman, R. 1999, \apj, 513, 879

\bibitem[Matthews et al. 2004]{jamie} Matthews, J.M., Kusching, R., Guenther, D.B., 
Walker, G.A.H., Moffat, A.F.J., Rucinski,
S.M., Sasselov, D., \& Weiss, W.W. 2004, Nature, 430, 51



















 

\bibitem[Richardson et al. 2007]{richard07} Richardson, L.J., Deming, D.,
Horning, K., Seager, S., \& Harrington, J. 2007, Nature, 445, 892  
 
\bibitem[Rowe et al. 2006]{rowe} Rowe, J.F., Matthews, J.M., Seager, S., Kuschnig, R., Guenther, D.B.,
Moffat, A.F.J., Rucinski, S.M., Sasselov, D., Walker, G.A.H., \& Weiss, W.W. 2006,
\apj, 645, 1241 

\bibitem[Rowe et al. 2007]{rowe07} Rowe, J.F., et al. 2007, submitted to \apj 
(arXiv:0711.4111) 








\bibitem[Seager \& L\'{o}pez-Morales 2007]{lopez} Seager, S. \& L\'{o}pez-Morales, M. 2007, \apj, 667, L191

\bibitem[Sharp \& Burrows 2007]{sharp07} Sharp, C.M. \& Burrows, A. 2007, \apjs, 168, 140 







\bibitem[Sudarsky \etal 2000]{Sudarsky00} Sudarsky, D., Burrows, A.,
\& Pinto, P. 2000, \apj, 538, 885


\bibitem[Sudarsky \etal 2005]{Sudarsky05} Sudarsky, D., Burrows, A.,
Hubeny, I., \& Li, A. 2005, \apj, 627, 520 

\bibitem[Swain et al. 2008]{swain} Swain, M., Bouwman, J., Akeson, R., 
Lawler, S. \& Beichman, C. 2008, \apj, 674, 482 (arXiv:astro-ph/0702593) 








\bibitem[van de Hulst 1974]{hulst2} van de Hulst, H.C. 1974, \aap, 35, 209













\end{thebibliography}
\end{document}